\documentclass[aps,prl,reprint,superscriptaddress]{revtex4-2}
\usepackage{amsmath}
\usepackage{amssymb}
\usepackage{graphicx, color}
\usepackage[all]{xy}
\usepackage{mathrsfs,amsmath,amssymb,amsthm,amsfonts,tikz,graphicx,accents,hyperref, color}
\usepackage{dsfont,epiolmec, latexsym, stmaryrd, comment}
\hypersetup{ linktoc=all,
    colorlinks, linkcolor={palatinateblue},
    citecolor={brightpink}, urlcolor={amaranth}
}

\graphicspath{{Images/}}

\definecolor{rosy}{RGB}{230,235,252}
\definecolor{myframetitle}{RGB}{90,89,170}
\definecolor{myblocktitle}{RGB}{140,185,249}
\definecolor{mytitle}{RGB}{10,80,26}

\definecolor{darkgreen}{RGB}{27,130,45}
\definecolor{darkblue}{rgb}{0,0,0.3}
\definecolor{darkred}{rgb}{0.7,0,0}

	\definecolor{light gray}{RGB}{220,220,220}
\definecolor{dark purple}{RGB}{108,0,217}
\definecolor{pink}{RGB}{190,20,100}
\definecolor{orang}{RGB}{193,63,0}
\definecolor{green}{RGB}{11,98,17}
\definecolor{darkpink}{RGB}{153,0,76}
\definecolor{bluegreen}{RGB}{0,102,102}
\definecolor{greenlagan}{RGB}{0,102,0}
\definecolor{redgreen}{RGB}{102,102,0}
\definecolor{Redgreen}{RGB}{153,76,0}
\definecolor{vividviolet}{rgb}{0.62, 0.0, 1.0}
\definecolor{amaranth}{rgb}{0.9, 0.17, 0.31}
\definecolor{palatinateblue}{rgb}{0.15, 0.23, 0.89}
\definecolor{brightpink}{rgb}{1.0, 0.0, 0.5}
\definecolor{cornflowerblue}{rgb}{0.39, 0.58, 0.93}
\definecolor{deepcarminepink}{rgb}{0.94, 0.19, 0.22}
\definecolor{radicalred}{rgb}{1.0, 0.21, 0.37}
\definecolor{darkmagenta}{rgb}{0.67, 0, 0.67}

\usepackage[most]{tcolorbox}

\tcbset{highlight math style={left=02mm,right=02mm,top=02mm,bottom=02mm}} 
\usepackage{empheq}

\usepackage{stackengine}

\begin{document}
\title{Revisiting bounds on neutrino dark matter interaction at spikes}	
\author{Yasaman Farzan}
\email{yasaman@theory.ipm.ac.ir}
\affiliation{School of Physics, Institute for Research in Fundamental Sciences (IPM),
P.O. Box 19395-5531, Tehran, Iran}

\begin{abstract}
Interactions between high energy neutrinos produced in active galactic nuclei (AGNs) and the dark matter (DM) particles in the dense spike around the supermassive black holes may lead to attenuation of the flux. This consideration along with the observation of a sizeable flux from at least four AGNs have been used to derive bounds  on the neutrino dark matter scattering cross section \cite{Cline:2022qld}. We show that the attenuation strongly depends on the details of the microphysics parameters. In particular if the mediator is a vector boson with a mass much smaller than the center of mass energy of the neutrino and dark matter, the energy loss of the neutrino and therefore the flux attenuation will be negligible so the upper bound on the cross section in this limit can be considerably relaxed.
		\end{abstract}
\maketitle

Supermassive black holes in the center of galaxies are expected to form a dense profile of dark matter in their vicinity known as spike 
\cite{Gondolo:1999ef,Shapiro:2022prq,Sharma:2025ynw,Vasiliev:2007vh,Shapiro:2016ypb}.  IceCube has identified at least four AGNs with supermassive black holes in the center as sources of high energy neutrinos \cite{IceCube:2018cha,IceCube:2022der,IceCube:2021slf,IceCube:2022der,IceCube:2024ayt}.  These neutrinos pass through
the dense spike. If they can scatter off the dark matter particles, they will lose their energy.  The evolution of the total flux (integrated over angles) is then given by the so-called cascade equation  \cite{Cline:2022qld} as follows 
\begin{equation} 
	\label{cascade} \frac{m_{DM}}{\rho_{DM}} \frac{d \Phi (E_\nu)}{dr}=-\sigma \Phi(E_\nu)+\int_{E_\nu}^{\infty}dE_\nu^\prime \frac{d\sigma (E_\nu^\prime \to E_\nu)}{dE_\nu}\Phi(E_\nu^\prime)
\end{equation}
in which $\rho_{DM}$ and $m_{DM}$ are the DM density and mass. The scattering cross section is denoted by $\sigma$. The mechanism(s) for the neutrino production yield a monotonically falling energy spectrum typically described by a power law $E_\nu^{-\gamma_\nu}$ with $\gamma_\nu>0$. Since $\Phi(E_\nu^\prime)< \Phi(E_\nu)$ in the second term of Eq. (\ref{cascade}), $$\int_{E_\nu}^{\infty}dE_\nu^\prime \frac{d\sigma (E_\nu^\prime \to E_\nu)}{dE_\nu}\Phi(E_\nu^\prime)< \sigma \Phi(E_\nu)\ ,$$
which means $d\Phi/dr$ is negative, implying attenuation \cite{Mondol:2025uuw,Cline:2022qld,Cline:2023tkp,Dixit:2026zkv}. In the extreme case of zero recoil energy, ({\it i.e.,} $\frac{d \sigma}{dE_\nu}\propto \delta (E_\nu -E_\nu^\prime)$), the right-hand side of Eq. (\ref{cascade}) vanishes  and no attenuation can take place. Taking the mediator of the scattering to be a vector boson, $Z^\prime$,  with couplings of $g_\nu$ and $g_{DM}$ to respectively  neutrinos and DM, we can write
\begin{equation}
\label{total-sig} \sigma=\frac{g_\nu^2 g_{DM}^2}{4\pi m_{Z^\prime}^2}\left[ 1- \frac{m_{Z^\prime}^2}{s}\log(1+\frac{s}{m_{Z^\prime}^2})\right] \end{equation}
where $s=2E_\nu m_{DM}+m_{DM}^2$ and
\begin{equation}
	\frac{d \sigma (E_\nu^\prime \to E_\nu)}{dE_\nu} =\frac{g_\nu^2 g_{DM}^2 }{4\pi} \frac{m_{DM}[1+(E_\nu /E_\nu^\prime)^2]}{(m_{Z^\prime}^2 +2 m_{DM}(E^\prime_\nu-E_\nu))^2}\ . \label{cross-diff}
\end{equation}
If  $m_{Z^\prime}$ is
much lighter than $\sqrt{2 E_\nu m_{DM}}$, $\sigma$ turns out to be energy independent   \cite{Cline:2022qld}.   Neglecting  the second term in Eq. (\ref{cascade}) in  this limit, Ref. \cite{Cline:2022qld}  finds $\Phi(E_\nu)=\Phi_0 \exp (-\sigma \int \rho_{DM} dr/m_{DM})$ (see endnote [102] of \cite{Cline:2022qld}) and  assuming a value for the initial flux, $\Phi_0$, uses the measured flux to constrain the attenuation and therefore $\sigma$ (see also \cite{Cline:2023tkp,Dixit:2026zkv}). The main point of the present study is that for $m_{Z^\prime}^2/(2 m_{DM}) \ll E_\nu$, the differential cross section is narrowly peaked at $E^\prime_\nu \simeq E_\nu$ so the second term in Eq.~(\ref{cascade}) significantly cancels the first term.

Let us define
\begin{equation}
	f(E_\nu,\frac{m_{Z^\prime}^2}{m_{DM}})=1-\frac{\int_{E_\nu}^{\infty}dE_\nu^\prime \frac{d\sigma (E_\nu^\prime \to E_\nu)}{dE_\nu}\Phi(E_\nu^\prime)}{\sigma \Phi(E_\nu)}
\end{equation}
so that the cascade equation can be rewritten as 
$$  \frac{m_{DM}}{\rho_{DM}}\frac{d \Phi (E_\nu)}{dr}=-\sigma f \Phi(E_\nu)\ .$$ For $f\ll 1$, the bounds on constant $\sigma$ obtained in \cite{Cline:2022qld} will be relaxed roughly by $1/f$.
Taking $\Phi$ to be  any decreasing function of energy, $f$  will be  a monotonically decreasing function of $E_\nu$. Moreover, as $m_{Z^\prime}^2/(E_\nu m_{DM}) \to 0$, $f$  goes to zero, too. 
With $\Phi \propto E_\nu^{-3.2}$  \cite{IceCube:2022der} and $m_{Z^\prime}^2/m_{DM}=2$~GeV (10~GeV),
we find that  $2\times 10^{-3}< f< 10^{-2}$ ($2\times 10^{-2}< f< 10^{-1}$) for energies relevant for the ICECUBE observation of NGC 1068 {\it i.e.,} for  $1.5~{\rm TeV}<E_\nu <15$ TeV. Such small values of $f$    dramatically relax the bound from NGC 1068 \cite{Cline:2023tkp,Dixit:2026zkv} which was considered to be the most stringent one \cite{Dixit:2026zkv}. In addition to attenuating the flux, large values of $\sigma$ ({\it i.e.,} $\sigma \geq m_{DM}/[f(1.5~{\rm TeV})\int \rho_{DM} dr]$ make the energy spectrum harder. Considering the uncertainties in the flux normalization prediction \cite{Fiorillo:2023dts,Inoue:2019yfs,Murase:2019vdl,Kheirandish:2021wkm,Murase:2022dog,Eichmann:2022lxh,Inoue:2022yak,Yoast-Hull:2013qfa,Fiorillo:2024akm}, the energy spectrum may be considered a complementary tool to constrain $\sigma$ even in the regime that it is constant in energy.

Even for the ranges of parameters for which $f$ is not very small, the most stringent bounds on $\sigma$ found in \cite{Cline:2022qld,Cline:2023tkp,Dixit:2026zkv} may be avoided by the following considerations: (1)  If the dark matter is symmetric, the couplings responsible for scattering can also lead to the pair annihilation of DM particles in the inner part of the spike to $\nu \bar{\nu}$ so the most stringent bounds coming from a spike density profile increasing down to the neutrino production region (BM1 and BM1' in \cite{Cline:2022qld,Cline:2023tkp,Dixit:2026zkv}) can be  avoided. As discussed in \cite{We}, avoiding such an annihilation is not trivial (perhaps impossible) even within the coannihilation scenario. (2) Within asymmetric dark matter model, annihilation cannot take place but neutrinos obtain effective mass inside the spike. As shown in \cite{We}, if  only $\nu_\tau$  couples to $Z^\prime$, $\nu_e$ and $\nu_\mu$ cannot oscillate to $\nu_\tau$ because of the suppressed effective mixing within the spike so the attenuation will be negligible despite significant impact on the final flavor ratio. If, in addition to $\nu_\tau$, $\nu_e$ ($\nu_\mu$) couples to $Z'$, attenuation will be at most $2/3$ ($1/3$) \cite{We} still tolerable within the uncertainties of the flux prediction \cite{Fiorillo:2023dts,Inoue:2019yfs,Fiorillo:2024akm,Murase:2019vdl,Kheirandish:2021wkm,Murase:2022dog,Eichmann:2022lxh,Inoue:2022yak,Yoast-Hull:2013qfa}.  {However, as pointed out in \cite{Cline:2023tkp}, if $Z^\prime$ couples equally  to all three neutrino flavors, there will be no matter effect to suppress the flavor oscillation.}

In conclusion,  generalizing the bounds  found in \cite{Cline:2022qld} to all parts of parameter space and to different dark matter models should be done with special care as the details of model can completely change the picture. There is still room for IceCube-Gen2 and KM3NeT to surprise us by discovering effects   induced  by neutrino  scattering off the spike as demonstrated in \cite{We}.

\textbf{Note added:} After the submission of the present article to the archive,  Ref. \cite{Cline:2026stf} appeared  in response to the present letter. The main disagreement between this letter and Ref. \cite{Cline:2026stf}  is their claim that the  cross section shown in  Eq.~(\ref{total-sig}), coming from the alleged  interference between  $t$- and $u$-channel diagrams, decreases with energy. However, this cross section comes only from  a $t$-channel contribution. Moreover, for  $E_\nu \to \infty$, it approaches a constant [{\it i.e.,} as $E_\nu \to \infty$, $\sigma \to g_\nu^2g_{DM}^2/(4 \pi m_{Z'}^2)$] rather than decreasing with energy. Indeed,  Eq. (\ref{total-sig}) of the present letter exactly corresponds to Eq. (8) of \cite{Cline:2026stf} as it should. The main point of the present letter is that in the high energy limit where $\sigma (E_\nu)$ approaches a constant ({\it i.e.,} when $s\gg m_{Z'}^2$), $d\sigma(E_\nu^\prime \to E_\nu)/dE_\nu$ approaches to a finite nonzero value times  $\delta (E_\nu^\prime -E_\nu)$ so according to the cascade equation in Eq. (\ref{cascade}), the attenuation will be suppressed. In contrast to this consideration, Ref. \cite{Cline:2022qld}, according to  its endnote [102],  ignores the second term in the cascade equation to obtain the bound they show in their Fig.~1. Thus, our conclusion is that the bounds in Fig.~1 of Ref \cite{Cline:2022qld}   can be considerably relaxed. 
However, our results are consistent with Figs.~3 and 4 of Ref.~\cite{Cline:2022qld} as they allow the range $m_{DM} E_\nu/m_{Z'}^2 \stackrel{>}{\sim} 1$ with $E_\nu =290$ TeV (the energy of the TXS 0506+056 neutrino event studied in Ref. \cite{Cline:2022qld}), consistent with the consideration in the present letter. The numerator of the differential cross section shown in Eq. (\ref{cross-diff}) slightly differs from that of Eq. (9) in Ref. \cite{Cline:2022qld}. However, this does not change  the argument as both formulas at $E_\nu'\gg m_{Z'}^2/(2m_{DM})$ are sharply peaked at  $E_\nu^\prime \simeq E_\nu$.

\begin{acknowledgments}
I would like to thank to M. M. Sheikh-Jabbari, J. Cline  and S. Abbaslu for useful comments.
\end{acknowledgments}

\bibliographystyle{fullsort.bst}

\bibliography{arXivreferences}
\end{document}